# Channels as Enzymes: Oxymoron and Tautology

*Topical Review*

## R.S. Eisenberg

Department of Physiology, Rush Medical College, Chicago, lllinois 60612 USA

### Introduction

The reproductive, developmental, and chemical life of cells is performed by protein catalysts called enzymes. Groups of enzymes work together to control the replication and expression of genes, to control the synthesis and degradation of each chemical constituent of cells. The electrical life of cells is controlled by the cell membrane, in particular, by proteins embedded in the membrane, called channels that regulate the flow of current[1]. This paper explores the analogy between enzymes and channels, arguing that channels and enzymes function in similar ways and so should be questioned in similar ways. Indeed, I argue that channels can be viewed as enzymes, as catalysts for the flow of electric current, and that this perspective helps the membrane biologist in his daily work, the design and interpretation of experiments.

### Biological Role of Electricity

Most cells use electricity. Nerve cells in the brain use electrical signals to detect and analyze these very words. Nerve axons transmit the words by propagating waveforms of voltage called action potentials: similar action potentials trigger movement of skeletal muscle and coordinate contraction of the heart. Indeed, death is defined by the end of electrical activity of the heart or the nervous system, depending on where you live and die. Even epithelial cells transporting uncharged sugars generate electrical current.

Electricity plays a central role in cells and tissues just as it does in most of our technology because the components execute functions accurately, quickly, and flexibly in little space, using little power, particularly compared to systems based on water flow or diffusion of molecules.

The widespread importance of electricity comes as a surprise to some biologists, but I suspect this surprise reveals more about our education than it does about biology. Few biologists are taught the essential language of electronics and electricity. the Laplace transform and Maxwelløs equations; most of us arc only qualitatively familiar with the properties and advantages of the electric field. although its properties (and the Laplace transform)·are widely taught to engineering students in their first year of university.

(In a way, the study of bio-electricity has been like the study of genetics before Watson and Crick discovered the chemical nature of the gene: both have been isolated from the mainstream by their specialized language and techniques.)

Biologists arc taught the language of chemistry, and that is certainly appropriate given the role of DNA as the blueprint and proteins as the machines of life. The question is how can the language of biochemistry be used to describe the electrical properties of cells? The answer arises from the application of two new techniques that allow measurement of the electrical properties of individual channels. The reconstitution method (Miller, 1986) makes vesicles of more or less natural membranes and then fuses these vesicles to an artificial bilayer, arranging conditions so only a few channels function in the bilayer at one time. The patch-clamp method (Sakmann & Neher, 1983) isolates (electrically, mechanically, and chemically) a patch of membrane with a seal of gigohms resistance formed (by an unknown mechanism) between the lipid of the cell membrane and the glass of a pipette. The isolated patch of membrane often contains only one functional channel from which current is measured while the pipette voltage is clamped to a known value.

The language of biochemistry can describe the properties of these channel proteins, and so it describes how these protein molecules control current flow through membranes, and thus many electrical properties of cells. This essay compares the biochemical description of enzymes and the traditional physiological description of conductances and channels. An enormous amount of work has been done to understand the role of protein catalysts (i.e., enzymes) in the chemical life of cells. This work has been going on long enough, has described enough different enzymes from enough types of animals, and has been molded by sufficient technological revolutions that it has a tradition, a set of agreed upon questions that most biochemists ask whenever any enzyme is studied.

This essay tries to ask those questions of ionic channels by looking at channels as enzymes, hoping that view is apt and useful in designing experiments, as well as amusing.

---

[1] Some important currents do not move through channels:
  (1) fluxes driven by coupled transporters like the sodium pump and
  (2) capacitive currents flowing across myelin or membranes, during propagation of the nerve action potential.

**Enzymes**

We all know what an enzyme is. It is a catalyst, a protein that changes the rate of a chemical reaction without changing its final equilibrium, the eventual concentration of substrates and products. Enzymes accelerate the rate at which concentrations of substrates and products change; that is to say enzymes increase the flux of substrate and product; they make it easier for substrates to change into products. To put it more formally, enzymes convert substrate to product by using the energy of those reactants, without contributing energy themselves—they simply modify the transition from substrate to product and so control the rate of a reaction. Enzymes make the transition between chemical forms easier. They lower the activation energy, the height of barriers between substrates and products. That is to say, enzymes stabilize the transition state by lowering its free energy.

**Channels**

Channels also do not use energy directly and do not change the free energy of ions on one side of a membrane or the other. That is well known and its discovery was in fact a key historical step along the path to understanding the mechanism ('the ionic basis') of the action potential (Hodgkin, 1964; Hodgkin, 1977). The idea of an ionic conductance $g_K$ for example, was introduced to describe a membrane process that did not use energy directly but did control the permeation of potassium. Today, we recognize $g_K$ as a measure of the number of channel proteins through which an ion can move, proportional to the conductance of a single channel, to the total number of channels of that type, and to the probability that a channel is open.

Channels control the flux of ions across membranes, increasing the flux by many orders of magnitude. The free energy that drives the movement of ions is just the concentration and electrical gradient so nicely called the driving force by Hodgkin and Huxley (1952). What they called driving force is simply the difference in electrochemical potential or the difference in free energy (per mole) of an ion between one side of a membrane and the other. Channels modify the rate of movement of ions, increasing the flux by many orders of magnitude, by something like a factor of $10^{17}$ because the lipid acts as a barrier some 67 times the thermal energy, namely 67 kT (Hille (1984, p. 188) computes the flux through a channel; Honig, Hubbell and Fleweling (1986, p. 170), compute the flux through lipid). Channels make it easier for ions to move, but they do not change the equilibrium any more than catalysts do.

Channels modify the flux of ions the same way enzymes modify the flux of reactants—they stabilize the transition state between substrate and product, if we define the transition state of a channel as the state with an ion in the pore (Fig. 1). The channel protein stabilizes this state (compared to what would happen without the protein) because it provides polarization charge to neutralize the permanent charge of the permeating ion. The channel protein has a high dielectric constant in the wall of its pore and lowers potential barriers to ion movement across the membrane.

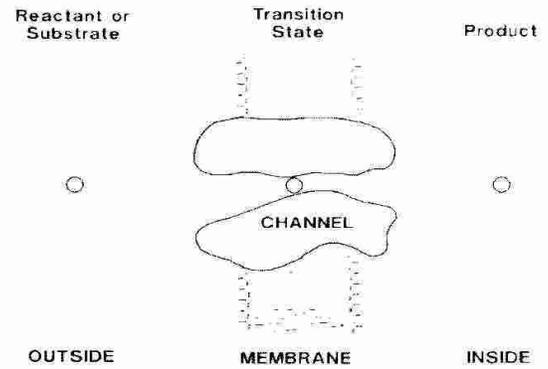

Fig. 1.

**Channels as Enzymes**

Channels then modify the rate of a chemical reaction just as enzymes do. And channels have substrates and products just as clearly defined as do enzymes. The substrate of a channel is just the permeable ion on one side of the membrane, and the product is just the permeable ion on the other side of the membrane. The substrate and product of channels have different free energies, just as they do for enzymes. Substrates and products of enzymes are different chemical species with different free energies at the same location. Substrates and products of channels are the same chemical species, at different free energies and different locations. The spatial gradient of electrochemical potential drives diffusion just as the chemical gradient of free energy drives a chemical reaction.

In this sense then a channel is a catalyst for diffusion through membranes and *a channel is an enzyme,* even if it does not mediate chemical reactions of the ordinary type.

**Who Cares if a Channel is an Enzyme?**

Whether or not a 'channel' is an 'enzyme' depends, of course, as much on the definition of those words as on the properties of those molecules. Textual analysis of the meaning of words is not one of the better ways to understand the world, particularly the world of science, and so there is an understandable lack of enthusiasm among scientists for questions like 'Is a channel an enzyme?' What a scientist really wants to know is 'What can we learn by considering channels as enzymes?' Or even more practically, 'Can we design better experiments or understand them more clearly by considering channels as enzymes?'

Looking at channels as enzymes is helpful because it links issues of ion permeation and channel gating to issues of protein structure, thus applying the insights of physical biochemistry and protein engineering to channology and *vice versa.* Physical measurements can tell us the static three dimensional structure of any protein that can be crystallized, a class that unfortunately does not yet include classical channel proteins, except melittin (Terwilliger & Eisenberg, 1982). Physical measurements can tell us a large amount about the rapid (nanosecond to picosecond) motions of spectroscopically observable parts of a protein. But physical measurements rarely focus on the part of the protein relevant to its natural function, whether that is catalysis or transport. Physical measurements do not often extend to the mechanistically relevant time scale of microseconds, let alone the physiologically relevant time scale of milliseconds to seconds. In short, what physical measurements tell of proteins is exquisite but perhaps not what we want to know. It is as if we knew all that could be known of a part of Michelangelo's *Pieta*, but of a part— say, the pedestal— that might not be of the greatest interest.

Channologists, on the other hand, study subjects of intrinsically great biological interest, namely the natural function of channel proteins (the control of current flow) and on the physiological time scale. But our resolution of measurement is appalling, with virtually no knowledge of three dimensional structure and no dynamic measurements available at all, unless one considers measurements of gating current to reflect the large persistent conformation changes associated with channel opening. The channologist can only see Michelangelo's *Pieta* myopically, at a distance. through a crowd; but he can watch in detail the reaction of viewers (which is the emotionally relevant output of the sculpture just as current is the biologically relevant output of a channel).

Viewing channels as enzymes then helps the channologist see his results in the image of real proteins. It helps him make models of channel proteins using elements and properties known to describe other better known proteins; it helps him avoid models built with mythological elements: rigid uncharged walls of pores are more implausible than unicorns; unicorns might exist, but channel proteins containing hundreds of amino acids but only a handful of dipoles or fixed charges cannot exist; motions of proteins taking microseconds cannot occur without billions of collisions and friction. On the other hand, viewing channels as enzymes may allow the enzymologist a closer approach to the natural function of his protein, on the physiological time scale, avoiding elaborate analysis of protein motions irrelevant to biological functions.

**Proteins are Complex**

Proteins are enormous objects on an atomic scale, capable of an incredible number of motions, possessing a larger than astronomical number of energy levels and conformations. Physical properties of the protein not involved in its biological function (e.g., its absorption of infrared radiation) can involve any or many of these conformations; indeed, the regions and conformations of the protein determining a physical property may change as conditions (i.e., ionic strength, pH, etc.) change. Thus, there is no reason to believe that any reasonably simple model can describe a general physical property of a protein over a reasonably wide range of ionic and biological conditions, particularly if the property is irrelevant to its function.

```
              ENZYMES
       Malcolm Dixon, Edwin C. Webb
             Table of Contents

PREFACE                                    xv
SYMBOLS AND ABBREVIATIONS                  xix
  I   INTRODUCTION                          1
  II  ENZYME TECHNIQUES                     7
  III ENZYME ISOLATION                     23
  IV  ENZYME KINETICS                      47
  V   ENZYME CLASSIFICATION               207
  VI  ENZYME SPECIFICITY                  231
  VII ENZYME MECHANISMS                   271
  VIII ENZYME INHIBITION AND ACTIVATION   332
  IX  ENZYME COFACTORS                    468
  X   ENZYME STRUCTURE                    519
  XI  ENZYME BIOSYNTHESIS                 570
  XII ENZYME BIOLOGY                      622
'ATLAS' OF CRYSTALLINE ENZYMES            664
TABLE OF ENZYMES                          683
REFERENCES                                974
INDEX                                    1077
```

Fig. 2.

**Protein Properties Selected by Evolution may be Simple**

Properties of proteins directly relevant to biological function may be much simpler to model and understand. Evolution may have selected a simple mechanism, like opening a pore, or lowering one energy barrier by changing the charge on one dipole, to perform an important physiological function, like controlling the flow of current into a cell. Thus, one has a greater hope of understanding a physical process involved in natural function than of understanding an arbitrary property of a protein. It may be possible to guess and test (i.e., scientifically determine) a unique and simple model that corresponds to the actual mechanism evolved by natural selection to perform the protein's function.

**Open Channel Permeation**

For this reason, it may be possible to construct a general theory of ion permeation *through an already open channel.* The structure involved is just the pore of the protein, and this is clearly much simpler than the overall structure of a protein. And the interactions of the wall of the pore with the permeating ion may well be as simple as the idealized interactions of an ion with a solvent, describable to a first order as a particle moving with friction through a potential.

An equivalent theory of classical enzymes seems far away. After all, a general theory of or ganic chemical reactions is not on the horizon, let alone at hand, even at biological temperature and physiological conditions. Similarly, a general theory of protein conformation change is not available, although we are so embedded in this field that we cannot glimpse a horizon, let alone guess what is over it.

In this very specific way, viewing channels as enzymes may help in understanding the properties of both. In another more general way, viewing channels as enzymes is quite helpful. Choosing questions and designing experiments depend as much on the sociology and psychology of scientists as they do on the logic of science. It seems worthwhile to ask questions of channels like those that have been useful to enzymologists, emphasizing those that have been productive and avoiding those that have been unproductive. Many more workers have studied enzymes for many more years than have studied ion permeation, at least with molecular resolution. A great deal more work has been done, and a great deal more is known; and more mistakes have been made and false trails followed, if only because so much more has been done. Knowing some of this history, we may be able to investigate channels more efficiently if channologists make our plans with conscious knowledge of the history and themes of our cousins, the enzymologists.

A classical place to start examining the themes of enzymology is the Table of Contents of a classic reference on enzymes, like *Enzymes* (Dixon & Webb, 1979; Fig. 2) that summarizes the classical (pre-recombinant DNA) knowledge of proteins.

**Isolation of Enzymes and Channels**

The title of Chapter 3, *Enzyme Isolation,* quickly reminds us of the *sine qua non* of protein chemistry: a preparation of enzyme has to be pure before it can

*Intentionally left blank*

be studied chemically. For many decades there were long arguments about the properties of enzymes, arguments caused, as we now know looking with hindsight, by impurities in the preparation. If a solution contains an unknown mixture of catalysts, it will be difficult to know what reactions are being catalyzed; it will be more or less impossible to know how the enzymes are influencing the reactions, what their mechanism of reaction is. An admonition is now accepted, õDon't waste clean thoughts on dirty enzymesö (Racker/Kornberg: see Kennedy, 1989).

What does enzyme purification have to do with channels? Consider a macroscopic piece of membrane. Isn't it just an impure mixture of proteins, in the sense that it contains several, perhaps many, channel types, which must be identified or separated before they can be understood?

I would argue that the isolation of channels is quite as important to membrane biology as the purification of proteins was to enzymology and for the same reasons. The corollary to the argument is that measurements from macroscopic membranes are as ambiguous and hard to interpret as those from unknown impure mixtures of enzymes. Fortunately, we now have the tools to separate and isolate channels easily with the gigaseal method of patch clamp and the reconstitution method of membrane biochemistry, and so membrane biology has enjoyed an extraordinary growth of knowledge and popularity in the last decade, as direct experimentation replaced indirect argument.

### Isolation of Channels in the History of Physiology

It is interesting to look at the history of electrophysiology in the context of this discussion, to cast a glance backwards to the study of macroscopic preparations looking for the importance of purity in channel preparations. It seems to me that the classical preparations of electrophysiology were all chosen for their purity, whether this choice was made consciously or not. (Purity was also sought in the strictly technical electrical sense: the preparation had to be pure enough, without series resistance, with simple enough geometry to permit voltage clamping.) And it seems to me that the most damaging problems in traditional electrophysiology, in traditional voltage-clamp experiments, reflected the unknown heterogeneity of conductances, even more than the somewhat known technical failures in the spatial and temporal control of voltage and concentration (e.g., Levis, Mathias & Eisenberg, 1983). Squid axon and frog neuromuscular junction have relatively simple structure and so pose fewer technical problems of voltage control; their membranes also contain only a few channel types. Squid membrane has just two channel types under most conditions, voltage activated Na and K channels. A third channel type is evident in decaying axons, a channel type historically called ÷leakageø, probably representing some sort of $Ca^{2+}$ activated nonselective cation channel. The frog neuromuscular junction has a number of channel types, but only one is activated by acetylcholine; thus, by studying just the agonist-activated conductance, a pure channel type could be isolated.

Despite the relative purity of these preparations, the speed of research was, in fact, limited by their impurity, particularly in preparations involving more than two channel types. Separation of even two channel types is not without its ambiguities, and differences in the results of separation can lead to different properties ascribed to each channel type. In many other preparations, containing more types of channels, unique interpretation of results proved impossible. The ambiguities associated with a large and unknown population of channel types are apparent in the history of cardiac electrophysiology, where many labs identified many channel types, not too many of which correspond to single channel currents recorded from these preparations.

Now we do not have to struggle so hard to find preparations dominated by just one or two easily controllable channels; now we can isolate channels directly by reconstitution or patch clamp. A channel reconstituted into a bilayer (by whatever means) is easily recognized in the ideal case. A gigaseal (formed by who knows what physics) isolates channels, electrically and functionally, making it easy to tell one from another in the ideal case. The historical role of the gigaseal in electrophysiology is rather like the role of the SDS gel in enzymology: each allows one experiment to do what had taken a career; each allows the easy purification and separation of types of proteins[2].

### Co-factors in Enzymes and Channels

It is well known, however, that enzymes must not be too pure if they are to function properly: many enzymes need *Enzyme Co-factors* (Ch. 9 of Dixon & Webb, 1979) to carry out their function. Some co-factors are small organic molecules like many vitamins, or even ATP; others are metal ions like $Ca^{2+}$. Some co-factors act as part of the active site of an enzyme, others stabilize or alter the conformation of the enzymes, some even are intermediates in the chemical reaction.

---
[2] Single channel recording, whether by reconstitution or patch clamp, is rarely as ideal as we have made it seem. Channels reconstituted into artificial lipid membranes will lose accessory proteins and may have modified properties associated with the preparative process. Channels isolated by gigaseals often have perplexing gating behavior, punctuated by inactivity, which is not easy to reconcile with known macroscopic properties of the channel's conductance. Indeed, the ability of single channel measurements to predict macroscopic measurements quantitatively has not been adequately explored in many channel types.

The analogy with channels seems so clear that it is hardly an analogy. Many channels have definite requirements for nonpermeant ions: cyclic AMP modifies the properties of $Ca^{2+}$ activated K+ channels, as does $Ca^{2+}$, of course. $Mg^{2+}$ modifies the properties of inward rectifiers. It is not yet clear to what extent these co-factors also control the function of the channel and act as allosteric effectors.

Most channels have definite requirements for $Ca^{2+}$ ions on one and the other side; most channels from the plasma membrane of cells function well if the solution on their outside contains a few mM $Ca^{2+}$ and the solution on their inside contains almost no $Ca^{2+}$, namely the sub-micromolar amounts usually found inside cells. Many channels require the presence of ATP and change their conformation if it is not present. Some channels change their properties dramatically in the absence of their substrate, Ca channels changing to K channels, for example, in the absence of $Ca^{2+}$. Here, the substrate is in a sense also a co-factor. Channel proteins cannot be too pure if they are to work physiologically, any more than enzymes can be.

### Naming Enzymes and Channels

Enzymes need names, even before they're fully purified, so Dixon and Webb (1979) spend many pages describing *Enzyme Classification.* The naming of enzymes became a serious problem as the number of enzymes grew, as isolation techniques became easier. Everyone started discovering and naming enzymes; everyone used different õstandardö conditions to assay their enzymes; and, of course, there were no shortage of different animals, bacteria, tissues, or cells as sources of these enzymes.

Isn't that where we are now in channology? õNewö channels are being reported at a wonderfully alarming rate, perhaps more than one a month, i.e., nearly one per issue of *J. Physiol. (London).* There is no standard nomenclature, no standard assay procedure, and, of course, no standard biological source for the molecules. So we often don't know which channel is which, let alone what they do. We need to standardize our nomenclature and assay conditions if we are to minimize confusion. Seasoned membrane biologists need to meet in Paris every year or two, following in the footsteps of enzymologists, set up international standards, and hopefully not change the name of everything we have already learned more than once or twice.

To begin discussion (but certainly not end it) I suggest a nomenclature emphasizing open channel properties, illustrated here for the acetylcholine (nicotinic) receptor

| *Agonist* | *Selectivity* | *Channel Conductance* |
|---|---|---|
| Acetylcholine | Cation | 40 pS |
|  |  |  |

*abbreviated to*
ACH-CAT-40

Another nomenclature, preferred by my colleagues in an informal survey, might emphasize gating properties, listing the agonist and selectivity, and the turn-on and turn-off mechanism, but not listing the single channel conductance because it is not diagnostic enough of a particular channel type. Many other possibilities exist that may be better than these: the point is one should be chosen. A common language like English, however, arbitrary its spelling, is better than no common language at all.

### Kinetics, Mechanism, and Blockers of Enzymes and Channels

Turning from nomenclature back to the Table of Contents of *Dixon and Webb,* we see chapters devoted to *Enzyme Kinetics, Enzyme Mechanisms* and *Enzyme Inhibition and Activation.* A substantial fraction of the literature on enzymes is devoted to studying the velocity of the catalyzed reaction (i.e., the flux of substrate into product) and how that depends on the concentration of substrate and enzyme, on pH, temperature, ionic strength and so on. This kinetic evidence is used to establish the sequence of sub-reactions that compose the overall enzymatic reaction, i.e., the enzymatic mechanism. The kinetic data describing the flux of the reaction is also used to evaluate the effects of inhibitors and activators on the reaction.

Some enzymes catalyze reactions at their active site without modifying their own conformation (these often fitting the Michaelis Menten formalism) whereas other enzymes clearly change their conformation, their alignment of polypeptide chains or subunits, while speeding a reaction. A major topic in each study is the division of effects into those involving a conformational change of the enzyme, called allosteric effects, and those involving reactions at a structurally fixed active site, reactions that might, however, be more complex than the traditional Michaelis-Menten scheme. Channels fit quite well into this scheme if one identifies the opening of a channel (usually called *gating)* with an allosteric conformational change (Catterall, 1977) and ion permeation with catalysis at an active site. In this view, the agonist is the allosteric effector and the open channel (i.e., the channel's pore) is the active site of the channel õenzyme.ö It is interesting that the separation between time-dependent properties of a channel conductance and voltage-dependent properties was clearly made by Hodgkin and Huxley (1952); indeed, to some extent by Cole (1947). The time-dependent properties (the so-called "instantaneous conductance") are now known to reflect the current-voltage relations of the (already) open channel, if the "instantaneous" measurement can be made in something less than say 50 μsec. The time-dependent properties (described by the evolutions of *m, h,* and *n* in the Hodgkin-Huxley formalism) are now known to reflect both the time and voltage-dependent properties of the gating process, the conformational change that opens the channel. Thus, in a certain sense, channologists anticipated the ideas of allostery by a few years.

Enzyme kinetics extend naturally into Dixon and Webb's (1979) chapters on *Enzyme inhibition and activation,* the study of the action of various agents on the velocity of reactions. There is no shortage of compounds that block or activate enzymes, and the task of determining if the blockage is noncompetitive or competitive (i.e., block at the same or different sites) has occupied many scientists for many years. One problem is particularly worth mentioning: enzyme inhibitors can bind at sites away from the active site and still have the kinetics of competitive inhibition, particularly if the binding of the inhibitor allosterically modifies the binding of the substrate and *vice versa.* If these two sites interact reciprocally (as if they were coupled by a rigid helix of protein acting like a child's seesaw or the connecting rod of a gasoline engine), the inhibitor and substrate seem to compete for the same site, although they physically bind at distinct locations.

The equivalent phenomena in channels involve the processes of channel activation and blockade. Chemical activators, usually called "agonists," are often the physiological regulators of channel opening and are thought in many cases to bind to sites outside the pore, although the evidence for this thought is not as direct as it might be. The voltage across the membrane is a common and important activator of some channels; the site of the voltage sensor is not so universally agreed. Some workers think the sensor is in the wall of the pore, sensing the potential within the pore itself; others think the sensor is near the lipid edge of the channel protein, away from the pore, sensing the voltage across the lipid part of the membrane. It is interesting that voltage-sensitive channel proteins have evidently evolved so only one group (the voltage sensor) produces a physiologically significant response to membrane potential: there are thousands of dipoles and charged groups in a channel protein, all of which must respond fairly dramatically to changes in the potential across the membrane, which, after all, involve changes in field strength of (at least) some $10 \text{ mV}/(5\times10^{-7}\text{cm})$ which is not a weak electric field, $20{,}000 \text{ V/cm}$. But the motions of most of these dipoles and charged groups induced by a depolarization to threshold are evidently decoupled from the channel conductance; such motions do not seem to modify the opening of the channel, the structure of the pore, or the interactions of permeating ions with the channel.

Blockers of channel permeation are usually artificial substances introduced to produce interesting experimental effects, particularly competition between blocker and flux, and are widely thought to permeate the channel's pore and interact there, perhaps at a binding site. This idea has been widely accepted, mostly because the competition between blocker and permeant ion is usually voltage dependent. The voltage dependence might, however, arise in quite a different way. If the binding of an inhibitor outside the pore moved a charged group in the membrane's electric field (e.g., a dipole in the wall of the pore) and the coupling between the binding site (outside the pore) and the charged group within the pore were rigid, the binding constant (outside the pore) would be allosterically modified by events within the pore, and *vice versa.* A mechanism of this sort is, of course, more complex than simple open channel blockade, but simplicity is not always (or even often) the rule in biology and the philosophical principle of Occam's razor (i.e., that one should accept the simplest of theories that fit experimental data) often cuts one's throat in the biological sciences.

## Selectivity in Enzymes and Channels

The separation of properties of channels into gating and open channel permeation extends to one of the more significant and famous membrane properties, namely selectivity. The ability of membranes to select between ions that are chemically not too different (e.g., $K^+$ and $Na^+$) is critical to the life of cells: it permits membranes of animal cells to maintain constant volume without having to support much hydrostatic pressure, thereby allowing animal cells to exist without a plant cell's rigid retaining wall. In this way animal membranes allow motility, contain proteins and nucleic acids within the cell, while still allowing metabolites in and out of the cell. Selectivity is now known to be a property (for the most part) of the open channel: the open channel permits larger flows of some ions rather than others, while the opening process depends much less on the identity of the permeating ion, in most cases. Thus, selectivity is a property of the pore of a channel, much as some kinds of *Enzyme Specificity* (Chapter 6 of Dixon & Webb, 1979) are a property of the active site and not the conformational change of enzymes.

Some specificity in enzymes is supposed, however, to result from a conformation change, from the fit of the enzyme to the substrate induced by the substrate's presence (Koshland, 1959), with the specificity depending on the plasticity as well as the structure of the active site (Bone, Silen & Agard 1989). Induced mechanisms for selectivity seem not to have been suggested for channels but are possible, even likely. A channel may be something like a snake swallowing a rabbit; it may change shape significantly, stretching or shrinking while the ion passes through, and the fit of the ion may thus depend on the structural change induced by the ion. The pore may not have a definite invariant size, independent of the strength of the interactions between channel protein and permeating ion (McCleskey & Almers, 1985). In this view of things, channel permeation is not so rigidly separated from channel gating, because both involve significant conformational changes, albeit not necessarily the same ones. The change in conformation of the channel induced by the permeating ion may be as important as the original conformation of the channel itself.

Here is a case in which the analogy between enzymes and channels suggests a new idea, at least to one channologist (but see Ring & Sandblom, 1988), and an idea that can be partially tested. If gating and open channel permeation are related, they should vary in a qualitatively similar way with a range of experimental interventions. Thus, it would be interesting to check whether pharmacological agents, divalent ions, permeating species and so on have similar effects on open channel conductance and on gating. For example, do related agents have the same sequence of potency in their action on the open probability function and on single channel current voltage relations? A thorough analysis of open channel noise under a variety of conditions might show the relation of gating to permeation: the noise of the open channel is likely to depend on (i) gating motions of the channel protein, (ii) fluctuations in the interaction between permeating ion and protein, and (iii) fluctuations in the number of current carriers (i.e., shot noise).

**Biosynthesis & Biology of Enzymes and Channels**

The macroscopic physiological properties of membranes dependô as we have seenô on how channels open, how they behave when they are open. Macroscopic properties also depend on what type of channels are present, on how the channels signal each other, and on where the channels are located within the cell. In other words, membrane properties depend on the biosynthesis and biology of channels just as cell metabolism depends on *Enzyme Biosynthesis* and *Enzyme Biology* (Chapters 11 and 12 of Dixon & Webb, 1979). The mechanisms that regulate the synthesis of channels and their location in membranes are just beginning to be investigated, but even a superficial glance at the diverse properties of different membranes in different cells makes it clear that such mechanisms exist and are of the greatest importance. The type and location of membrane channels are just as characteristic of a cell as are the type of its enzymes.

The interactions of channels are also just beginning to be investigated, but even now it is clear that channels and membranes proteins interact with each other by passing chemical messages back and forth, for example, cyclic AMP, GTP, or inositol tris-phosphate, forming pathways of some intricacy, although perhaps not as intricate as those of intermediary metabolism. The incredibly complex, but specific and important, pathways of intermediary metabolism took many decades to discover, and it was perhaps just as well that the pioneers did not have a glimpse of the tangle of reactions that have always filled cells and now fill biochemistry textbooks. It may similarly be just as well that channologists working on channel interactions do not know yet how it will all come out. If the complexity of channel interactions proves to be anything like that of enzyme interactions, pioneers might be discouraged. Channologists; like most scientists, probably proceed best if they take one step at a time, watching their feet lest they stumble, keeping their heads out of the clouds, while they seek, step by step, their personal heavens of truth and beauty.

*Intentionally Left Blank*

**Structure of Enzymes & Channels**

The last chapter of *Enzymes* we consider is *Enzyme Structure,* Chapter 10. By now many enzymes have been crystallized, x-rayed, and their structures analyzed until the static position of every atom is known with frightening precision. Only one intrinsic membrane channel has been crystallized that I know of, and so channel structure is unknown in any general atomic sense. One should wonder what use is a discussion of channel structure without structural data.

I would argue that something very important is known about channel structure and that this knowledge allows some insight into the role of channel structure in channel function, into how channels act as catalysts for diffusion. Channels are known to contain pores that allow ions to move at much the same rate they move in solution. The energy barriers in the pores must be low, the interactions between channel and permeating ion must be weak, and the pores of ionic movement must be like that of ionic movement in solution.

**Ions Moving Through Already Open Channels**

Ions move in solution in response to concentration and potential gradients, their speed of motion being determined by that driving force and the retarding friction caused by interactions with the solute. The source of this friction *in* a polar solvent like water is not known for sure, but is likely to depend on dissipative interactions between water molecules induced by the movement of an ion (so called ÷dielectric frictionø) at least as much as on collisions between the ion and the solvent molecules. One can begin, in any case, to describe the movement of one ion in solution as the random diffusion of a particle in a potential field, a particle randomly wandering back and forth through a field of viscous mud, a field that biases the motion because it slopes up and down the side of a hill. The analogy here between a random walk through a muddy gravitational field and ionic diffusion through the electric field in a channel is rather precise.

In a similar spirit, we have argued (Cooper et al., 1988*a,b*), and we were certainly not the first (e.g., reviewed in Levitt, 1986; *see also* Cooper, Jakobsson & Wolynes, 1985; Jakobsson & Chiu, 1987), that the best starting model for ion permeation through an open channel is the Fokker-Planck equation, perhaps the simplest equation that describes a random walk, a diffusion controlled by a potential and a friction.

$$\frac{\partial}{\partial x} J(x,t) = -\frac{\partial}{\partial t} p(x,t) \quad (1)$$

$$J(x,t) = \mu(x) \cdot p(x,t) - \frac{\partial}{\partial x}\{D(x) \cdot p(x,t)\} \quad (2)$$

where $p(x,t)$ is the probability density function for the location of a single particle (units: cm$^{-1}$) at location $x$ and time $t$; $J(x,t)$ is a probability density function describing the flux of a single particle (units: sec$^{-1}$), $D(x)$ is the diffusion coefficient that describes frictional effects on the particle; and $\mu(x)$ is the drift (units: cm/sec) of the particle in a potential field $U(x)$ (units: V/cm).

$$\mu(x) = -(\text{mobility}) \cdot \frac{\partial U(x)}{\partial x} = -\frac{zD(x)}{kT}\frac{\partial U}{\partial x} \quad (3)$$

where we have glided by the sometimes confusing definition of mobility (defined properly in Bockris & Reddy, 1970, pp. 371 and 376-377) and used the Einstein expression for the coefficient of the gradient of potential in terms of the charge on the particle $z$ and the thermal energy $kT$.

The potential $U(x)$ reflects both the potential across the membrane and the conservative interactions (i.e., potential energy) of the ion, the solvent, and the protein. The friction (described here by the diffusion coefficient $D(x)$) reflects the dissipative (e.g., collisional) interactions of the ion, the solvent, and the protein. The entire interaction of the ion, solvent, and protein is captured, in this oversimplified model, by the potential and friction functions. Thus, in a very real sense the only structures relevant to ion permeation (in this model) are the structures of those functions, the spatial variation of potential and friction through the channeløs pore.

**What is Friction and Potential in a Pore?**

The meaning of friction and potential can perhaps be seen more clearly if one considers a hypothetical sinusoidal motion of an ion within the pore of a channel protein, a protein considered as a collection of charges, dipoles, etc., tied together by springs and dashspots. The protein is considered as a macroscopic object, modeled as a set of masses connected by elastic bonds (i.e., springs) that conserve energy and frictional restraining elements (dashspots) that dissipate energy into heat. The masses are continually perturbed by random thermal motion, the whole model being in the spirit of the Langevin equation of Brownian motion, central to most analysis (Arnold, 1973; Ch. 1; Gardiner, 1985, pp. 80-83) and simulation (Allen & Tidesley, 1987, Ch. 9) of stochastic diffusion. The only interaction of ion and protein is electrostatic in the oversimplified model of this paragraph, but inclusion of collisions does not change the treatment in any important way; it just adds another frictional term (Cooper et al., 1988). (In classical derivations of the Fokker-Planck equation $D(x)$ reflects only the collisional terms.) If the ion moves at a very high frequency, compared to the natural frequency of the springs and dashpots of the protein, the charges etc. in the protein do not have time to move in response to the ion's motion and so the ion interacts with the original electrostatic potential of the static protein structure, that is to say, the ion interacts with the electric field **E** computed from the original distribution of charge in the protein using Coulomb's law (in the form describing charges in a vacuum) to compute explicitly the interactions of the ion with all the charge in the protein. That is to say, none of the interactions are hidden, i.e., made implicit by the use of Coulomb's law with a dielectric constant I. (I should perhaps add here a note of skepticism about the validity, if not utility, of simulations of molecular dynamics in which charge interactions with protein are described implicitly by a dielectric constant, by a single real number, a dielectric constant independent of time, experimental conditions, or the velocity of charge movement. Proteins are characterized by incredibly complex interactions with the electric and electromagnetic field over the entire range of frequencies from DC to x-ray, and so it is unlikely that the electric force between an ion and a charged group can be characterized by a dielectric constant independent of time or frequency, velocity, or experimental conditions. Furthermore, experimental variables like flux typically depend exponentially on dielectric constants, so small errors in the constancy of the dielectric interactions produce large errors in the biologically relevant variables.)

Turning now to the other extreme of frequency, we can perform a similar analysis of ion and protein interactions. If the ion moves at a low frequency, compared to the motions within the protein, the charge distribution in the protein has time to respond to the motion of the ion and so the ion interacts with the so-called potential of mean force, exactly the potential used in Debye Hückel theory, a potential that arises from the charge distribution of the protein after it has adjusted to the ion's position. The electric field present after the charge in the protein has completed this adjustment is the **D** field introduced by Maxwell given by the low frequency dielectric constant times the electric field, remembering that the dielectric constant in the sinusoidal case is a complex, frequency-dependent number not usually equal to one. In the time domain the physical process relating **D** and **E** in a channel or membrane are the same, but the mathematics is much more complex and awkward, described by a convolution integral.

In the general case, the motion of the permeating ion will be at speeds comparable to a significant number of the motions induced in the protein. The motions in the protein will follow the motion of the ion with some lag, with some phase angle in the sinusoidal case. The induced motion can be resolved into in-phase and out-of-phase components (as can any sinusoidal motion and thus most any motion, using Fourier analysis), components often described by real and imaginary numbers, respectively. The in-phase component of induced motion represents a frictional interaction just as the in-phase component of a current in an electrical circuit represents the dissipative interaction of electrons moving through matter, energy lost to heat in the resistors of the circuit. The out-of-phase component represents the conservative interaction just as the out-of-phase component of current in an electrical circuit represents energy stored in a capacitor (i.e., electric field) or inductor (i.e., magnetic field). The frictional interactions within the protein involve the loss of energy to heat; this energy can only be supplied by the ion's motion, and so the internal friction of the protein becomes a friction "felt" by the permeating ion: the permeating ion supplies the energy lost to heat in the dissipative motions induced inside the protein.

In this way we can make a precise operational definition of friction and potential forces, following in the footsteps of many others, no doubt. For sinusoidal motion of an ion, friction accounts for the in-phase force acting on the ion and potential accounts for the out-of-phase forces on the ion. Both friction and potential are "effective" quantities that will depend on frequency in the sinusoidal domain and on time in the time domain. In the time-dependent case, these quantities depend on an integral over time and so show memory effects, but turning to the sinusoidal domain avoids this complexity and thus simplifies understanding: as long as the underlying differential equations are linear, no physics is lost by considering just the sinusoidal case.

The friction and potential effective for ionic motion can be viewed as the friction and potential effective for ions moving sinusoidally at average velocities close to thermal velocity. The potential has two components: one due to interactions with the channel protein and the other due to the potential applied experimentally across the membrane and channel. The thermal (i.e., *rms* average) velocity of an ion permeating a channel is hardly affected by the transmembrane voltage (i.e., drift induced by the applied field) at room temperatures, so the friction and interaction potential in a channel should not depend much on experimentally applied voltage or current, at least in this simple view of things.

**More Realistic Description of Open Channel Permeation**

If we view ionic permeation as a random walk through a potential field impeded by friction, the Fokker-Planck equation is the starting point for a theory, but it is not the ending point. Channology needs to reach well beyond the Fokker-Planck equation in two directions, towards both atomic and biological reality. Atomic reality requires that the structure of the protein be introduced into the Fokker-Planck theory: the potential and friction of the Fokker-Planck equation must be related to protein structure and dynamics. Biological reality requires that the theory be extended to describe the single filing behavior characteristic of real ionic channels.

The relation of the structure of the channel protein and the potential and friction has been briefly discussed already. The discussion implies that both the potential and the friction depend on the dynamics of the protein's internalmotions as well as on the static locations of the atoms of the proteins. Or to put the same thing in a more traditional jargon, the interactions of a permeating ion and the channel protein depend critically on the local dielectric constant, both its real and its imaginary parts, at each location in the channel's pore, including its frequency dependence. This local dielectric constant cannot be determined over the biologically relevant frequency domain by the techniques of x-ray crystallography or nuclear magnetic resonance, to the best of my knowledge, and so I think it impossible for models derived from those techniques to quantitatively predict ion permeation. (Analysis of a mechanical model of gramicidin leads to a similar conclusion: Roux & Karplux, 1988). Measurements of the static structure of proteins can give qualitatively useful information about the potential function, for example, the likely number and location of potential barriers and wells; perhaps measurements of current voltage relations of channels under a wide variety of conditions can determine the size of these barriers and wells, if those measurements are interpreted with a theory based on the Fokker-Planck equation using a potential function compatible with structural information. It is also possible that measurements of open channel noise, and open channel block (both of which probe the potential function in different ways) will help in this regard when interpreted with the proper generalization of the Fokker-Planck theory.

But the theory based on the Fokker-Planck equation must describe the single filing behavior of ions so prominent in most ionic channels. A Fokker-Planck equation must be written for two interacting ions and solved for the general case, including relatively low energy barriers. Levitt (1986) and Gates, Cooper & Eisenberg (1989) have introduced interactions by writing a state diagram for a channel, assuming that a channel can be occupied by not more than one ion at a time, because of electrostatic repulsion between ions. This approach is important because it includes the essential property of single filing; it is appealing because of its elegance and the simplicity of the resulting expressions for channel current, some of which are nearly identical to expressions from the Michaelis-Menten theory of enzyme reactions. But the range of validity of the one ion model will not be known until the two ion Fokker-Planck equation is properly analyzed. One hopes that such analysis will support the lovely and powerful results of the one-ion models and, in particular, will support and extend the flux expressions so closely related to enzymology.

Which is probably not a bad place to end this essay. Channels are in many ways analogous to enzymes. In a certain sense channels catalyze diffusion the way enzymes catalyze chemical reactions. The strategy and tactics of enzymology are useful when investigating channels. And a theory of channels produces expressions closely related to those of enzyme kinetics.

We might conclude then that

**Channels *are* enzymes,**
.
.
.
nearly.

Acknowledgement

It is a pleasure to thank Tom DeCoursey and Fred Quandt for sharing insights into channel behavior
and
**John Edsall for a lifetime of advice and help**
and many useful comments on this paper.